\documentclass{emulateapj}
\def\stacksymbols #1#2#3#4{\def\theguybelow{#2}
        \def\verticalposition{\lower#3pt}
        \def\spacingwithinsymbol{\baselineskip0pt\lineskip#4pt}
        \mathrel{\mathpalette\intermediary#1}}
\def\intermediary #1#2{\verticalposition\vbox{\spacingwithinsymbol
        \everycr={}\tabskip0pt
        \halign{$\mathsurround0pt#1\hfil##\hfil$\crcr#2\crcr
                \theguybelow\crcr}}}
\def\lta{\stacksymbols{<}{\sim}{2.5}{.2}}
\def\gta{\stacksymbols{>}{\sim}{3}{.5}}

\begin{document}

\title{CREATION OF THE X-RAY CAVITY JET AND 
ITS RADIO LOBE IN M87/VIRGO WITH COSMIC RAYS; 
RELEVANCE TO RELIC RADIO SOURCES}

\author{
William G. Mathews\footnotemark[1] and 
Fabrizio Brighenti{\footnotemark[1]$^,$\footnotemark[2]}
}

\footnotetext[1]{UCO/Lick Observatory,
Dept. of Astronomy and Astrophysics,
University of California, Santa Cruz, CA 95064}

\footnotetext[2]{Dipartimento di Astronomia,
Universit\`a di Bologna,
via Ranzani 1,
Bologna 40127, Italy}


\begin{abstract}
Young cavities in the X-ray emitting hot gas in galaxy clusters are 
often filled with radio synchrotron emission from cosmic rays.
However, in the M87/Virgo cluster, 
where cavities are less prominent,
X-ray observations show a 30-kpc long
nearly radial filament of relatively cooler gas
that projects from the cluster core into a large (40 kpc) 
radio lobe. 
We describe the dynamical relationship between these
two very dissimilar observations with 
gas dynamical calculations that include the 
dynamical effects and spatial diffusion of cosmic rays.
After cosmic rays inflate the cavity, 
they diffuse through the cavity walls, forming a much larger lobe. 
The cavities, which are most visible just after 
they have formed (in about 20 Myr), require a 
total cosmic ray energy that is more than 10
times larger than that usually assumed, $E = 4PV$.
During the relatively brief cavity lifetime, a jet-like, low-entropy
thermal filament is formed in the buoyant flow and moves at high
subsonic velocities through the cavity center and beyond.
After 100 million years, long after the cavity has disappeared, 
the relatively dense filament 
extends to 20-30 kpc and the cosmic rays have diffused
into a quasi-spherical lobe 40 kpc in diameter. 
These computed X-ray and radio features agree well with those 
observed in M87/Virgo and resemble those in other 
``relic'' cluster radio sources such as Abell 13 and Abell 133.
Eventually, the filament falls back and shocks at the 
center of the cluster -- perhaps stimulating 
the famous non-thermal M87 jet -- and only the aging 
radio lobe remains. 
\end{abstract}


\section{Introduction}

The means by which accretion energy is communicated from 
massive black holes in cluster-centered galaxies to 
the surrounding hot gas are commonly thought to be fundamentally 
non-thermal. 
In galaxy clusters non-thermal energy is evidently 
transported and deposited in distant cluster gas via jets  
visible at radio frequencies. 
This notion is supported by numerous observations 
of concentrated non-thermal radio emission in young 
kpc-sized cavities in the hot cluster gas observed with X-rays
(e.g. Bohringer et al. 1993; Fabian et al. 2000; 
Birzan et al. 2006; Clarke et al. 2006).
A full understanding of these remarkable energy outbursts
can only be achieved with quantitative observational and 
theoretical studies of both cosmic rays and hot thermal 
gas in galaxy clusters.

In Mathews \& Brighenti (2007) we showed how a localized 
source of cosmic rays, as for example the working surface of 
a jet, carves out 
cavities in an initially uniform hot gas. 
When cosmic rays diffuse rapidly through the thermal gas,
the cavities produced with a given total cosmic ray energy 
are small and tend to disappear 
rapidly after the cosmic ray source turns off.
If the same total energy is imparted to cosmic rays that
diffuse more slowly through the gas,
the size of the cavities and their longevity both increase. 
In this case the  
non-thermal energy is efficiently trapped within the cavity
(Churazov et al. 2001).
The case of most interest probably lies between these 
extremes. 
Non-thermal radio emission 
from cosmic ray electrons is largely confined 
within cavities (or jets) at early times, 
but at later times the cosmic rays diffuse into the 
surrounding hot gas to form 
radio lobes with sizes much larger than the cavities. 

In the exploratory calculations described here 
we extend our previous calculations to 
study the 2D evolution of cavities formed with cosmic rays 
in hot cluster gas confined by gravity.
Because of the diffusive nature of cosmic rays 
and the requirement that they eventually evolve into the large 
radio lobes observed, 
the total non-thermal energy required to create typical 
X-ray cavities of radius $r_{cav}$ can be much larger 
than the purely hydrodynamic work $\sim P 4 \pi r_{cav}^3/3$ 
required to displace uniform hot gas with pressure $P$.

Another, more specific motivation for the calculations described 
here are the {\it Chandra} X-ray observations of 
M87/Virgo by Young et al. (2002) and Forman et al. (2005; 2007)
that reveal a $\sim25$-kpc long, nearly radial  
gaseous filament extending from 
the central galaxy M87 toward the SW.
(We adopt a distance of 16 Mpc from Tonry et al. 2001.)
This thermal feature, 
beautifully shown in Fig. 2 of Forman et al. (2007), 
is remarkable because it lies just along the radius of 
the large SW quasi-spherical ``outer'' radio lobe 
of radius $\sim25$ kpc observed at 90 cm
with the VLA (Owen et al. 2000). 
This geometrical relationship between the thermal filament and 
radio lobe can be seen in Fig. 5 of Young et al. (2002). 
By including the dynamical and diffusive effects of cosmic rays, 
we show how X-ray cavities evolve into radial filaments 
surrounded by large (radio) lobes. 
Our calculation described here is also relevant the 
evolution of some ``relic'' radio sources observed in other clusters. 
According to Giovannini \& Feretti (2004) these extended, 
steep spectrum radio sources are usually 
located in the outer parts of the hot cluster gas, 
unrelated to any non-central galaxy. 
However, cosmic rays that  
diffuse from X-ray cavities can only produce relic radio lobes 
that are relatively close to the cluster core, such as those 
in Virgo, A13 or A133. 
Most relic sources are too large 
(400 - 1500 kpc) or too distant ($\gta 1000$ kpc) to be 
explained in this way.

Previous computational studies of the evolution of X-ray cavities, 
in which cavities filled with 
super-heated, non-relativistic buoyant gas are formed or placed 
in the hot gas cluster atmospheres, 
have shown that the 
cold gas flowing around and beneath the buoyant cavity 
eventually projects up through the cavity, resulting in a 
radially outflowing thermal feature that is cooler 
and more metal-rich than the surrounding gas
(e.g. Reynolds et al. 2005; Gardini 2007; 
Ruszkowski et al. 2007; Roediger et al. 2007).
We therefore envisage a cavity-lobe evolution with two jets.
The first jet is an AGN jet, perhaps similar to
the famous 2 kpc non-thermal jet in M87, 
that injects cosmic ray energy into expanding cavities.
While it is commonly (even universally) 
assumed that X-ray cavities are formed by jets from the galactic 
nucleus, there is very little observational evidence 
of these jets. 
The second jet is the thermal jet that forms 
from the vortex flow around the buoyant cavity which, 
as we show here, can project to great radial distances.
We shall refer to this as a ``cavity jet'' because of 
its origin and resemblance to the long radial filament visible in 
{\it Chandra} images. 
We show below that the cavity jet can be visible long 
after the associated X-ray cavity has disappeared from view 
and the cosmic rays that initially created it have 
diffused into a large (radio) lobe surrounding the 
cavity jet.

In our discussion of the cavity-lobe evolution that follows 
we have elected not to compute the radio frequency 
surface brightness in detail although this will be essential 
to fully confirm the accuracy of our proposed evolution. 
We impose this limitation to avoid introducing 
new uncertain parameters -- such as the radial variation 
of magnetic field strength in the cluster gas, 
the ratio of of cosmic ray electrons to protons, 
synchrotron losses, etc. -- that would further complicate 
the dynamical presentation which is sufficiently 
detailed already. 
Consequently, we shall regard our model for the radio lobe 
a success if the cosmic rays occupy approximately 
the same volume as the outer radio lobe in M87/Virgo 
and have enough energy to produce the radio luminosity 
observed. 
Finally, our intention is to study the physical nature 
of the cavity-radio lobe evolution and not attempt to 
solve the global cooling flow problem 
although the two are obviously related. 

\section{Equations and Computational Procedure}

The combined Eulerian evolution of (relativistic) 
cosmic rays (CRs)
and thermal gas can be described with 
the following four equations:
\begin{equation}
{ \partial \rho \over \partial t} 
+ {\bf \nabla}\cdot\rho{\bf u} = 0
\end{equation}
\begin{equation}
\rho \left( { \partial {\bf u} \over \partial t}
+ ({\bf u \cdot \nabla}){\bf u}\right) = 
- {\bf \nabla}(P + P_c) - \rho {\bf g}
\end{equation}
\begin{equation}
{\partial e \over \partial t}
+ {\bf \nabla \cdot u}e = - P({\bf \nabla\cdot u}) 
- (\rho/m_p)^2 \Lambda
\end{equation}
\begin{equation}
{\partial e_c \over \partial t}
+ {\bf \nabla \cdot u}e_c = - P_c({\bf \nabla\cdot u})
+ {\bf \nabla\cdot}(\kappa{\bf \nabla}e_c) 
+ {\dot S}_c
\end{equation}
\begin{equation}
{ \partial \rho z_{Fe} \over \partial t}
+ {\bf \nabla}\cdot\rho z_{Fe} {\bf u} = 0
\end{equation}
where we suppress artificial viscosity terms.
Pressures and thermal energy densities in the plasma
and cosmic rays are related respectively by
$P = (\gamma -1)e$ and $P_c = (\gamma_c - 1)e_c$
where we assume $\gamma  = 5/3$ and $\gamma_c = 4/3$.
The cosmic ray dynamics are described by
$e_c$, the integrated energy density over the cosmic ray
energy or momentum distribution,
$e_c \propto \int EN(E) dE \propto \int p^4 f(p)(1+p^2)^{-1/2} dp$.

The first three equations are the usual equations for 
conservation of mass, momentum and energy in 
the hot thermal cluster gas, except we allow for 
optically thin radiative losses with the usual 
bolometric cooling coefficient 
$\Lambda(T,z)$ erg cm$^3$ s$^{-1}$ as described by 
Sutherland and Dopita (1993).
Note that the CR pressure gradient in equation 2 
contributes to the motion of the thermal gas. 
This exchange of momentum between CRs and gas arises
as the CRs 
diffuse through magnetic irregularities (Alfven waves) 
that are nearly frozen into the hot thermal gas.
However, magnetic terms do not explicitly enter in 
the equations because typical magnetic fields in cluster 
gas $\sim 1-10\mu$G (Govoni \& Feretti 2004) 
are too small, i.e.  
the magnetic energy densities $\sim B^2/8\pi \lta 10^{-11}$
erg cm$^{-3}$ are generally much less than the thermal
energy density in the hot gas. 
In addition, the Alfven velocity
$v_A = B/(4\pi \rho)^{1/2} = 2 n_e^{-1/2} B(\mu{\rm G})$
km s$^{-1}$ is typically much less than
the sound or flow speeds in cluster gas 
so the Alfven velocity of the magnetic scatterers can be 
ignored (e.g. Drury \& Falle 1986, Jones \& Kang, 1990).
Finally, the gas pressure will be isotropic, as we 
assume, if the scale of magnetic 
field irregularities is small compared to length 
scales of astronomical interest.

Although hot cluster gas is weakly magnetic, 
$\beta = 8 \pi P/B^2 > 1$, 
the proton Larmor radius 
$r_L = (m_pc/eB)(2kT/m_p)^{1/2} 
= 4 \times 10^8 T_7^{1/2}/B_{-5}$ cm
is about eleven orders of magnitude smaller than the Coulomb 
mean free path
$\lambda = 3 (3/\pi)^{1/2} (kT)^2 / 4 n_e e^4 \ln \Lambda
= 25 T_7^2/n_{e-2}$ pc, 
where numerical subscripts refer to exponents in cgs units. 
Note that $\lambda$ is 
much less 
than the size of most 
X-ray cavities.
We therefore assume that 
the dissipationless gas dynamics equations 
above are appropriate for hot cluster gas 
and that dissipation in shocks can 
be treated in the normal way with artificial viscosity.
The physical nature of dissipation in cluster 
gas is currently unclear (e.g. Lyutikov 2007).

Equation 4 above describes both the advection of CRs 
with the gas and their diffusion through the gas.
A mass conservation equation for the CRs is unnecessary
because of their negligible rest mass.
The CR diffusion coefficient $\kappa$ 
is difficult or impossible to calculate in the absence 
of detailed information about the magnetic field topology 
which is currently unknown. 
We expect that $\kappa$ may vary inversely with the 
density of the thermal gas, assuming that the magnetic 
field strength also scales with density.
For simplicity we ignore for now any dependence of $\kappa$ on 
CR particle momentum.
Since observed radio lobes are
approximately spherical rather than systematically oblate or prolate,
we assume that $\kappa$ is isotropic, which is
consistent with a highly irregular magnetic field on scales of
interest here.
For these preliminary calculations it is not necessary
to specify the CR composition, either electrons or protons
can dominate as long as they are relativistic.
Finally, we assume that the total CR energy density is not 
substantially reduced by losses due to synchrotron emission 
or interactions with ambient photons or thermal particles 
during the cavity evolution time. 
We show below that this assumption is appropriate for 
the Virgo cluster.

The set of equations above, including equation (5) for 
the conservation of iron, are solved 
in $z,r$ cylindrical coordinates using a substantially 
modified version of ZEUS 2D (Stone \& Norman 1992). 
The computational grid consists of 200 equally spaced 
zones in both coordinates out to 50 kpc plus an additional 
100 zones in both coordinates that increase in size 
logarithmally out to $\sim2$ Mpc.
The initial radial variation of 
gas density and temperature in M87 and the surrounding Virgo 
galaxy cluster are taken from the 
analytic fits to the observations suggested by 
Ghizzardi et al. (2004). 
We choose a spherical gravitational field 
${\bf g} = (g_z,g_r)$ that establishes exact hydrodynamic 
equilibrium for the corresponding (observed) 
gas pressure gradient.
The initial abundance profile in the hot gas is that 
suggested by Rebusco et al. (2006). 
The cosmic ray diffusion term in equation 4 is solved 
using explicit differencing so each time step is restricted 
by the stability condition for the diffusion as well 
as the Courant condition for hydrodynamic stability. 

We assume that the X-ray cavity is formed at some fixed
radius by CRs supplied or produced by a non-thermal jet 
from the central black hole (AGN). 
The CRs are deposited in a gaussian-shaped 
sphere of characteristic 
radius $r_s = 2$ kpc located at 
${\bf r}_{cav} = (0,10{\rm kpc})$, 
i.e. 10 kpc along the $z$-axis.
The CR source term in equation 4 is therefore
\begin{equation}
{\dot S}_c = {E_{ctot} \over t_{cav}}
{e^{-(({\bf r - r}_{cav})/r_s)^2} \over \pi^{3/2} {r_s}^3} 
~~~{\rm erg}~{\rm  cm}^{-3}{\rm s}^{-1}
\end{equation}
when $t < t_{cav}$.
The integral of $ (r_s \pi^{1/2})^{-3}e^{-(r/r_s)^2}$ 
over space is unity.
In the course of these calculations we have considered 
several values of the total CR energy $E_{ctot}$ and 
injection time $t_{cav}$, 
but we begin by describing only the flow that develops when 
$E_{ctot} = 1 \times 10^{59}$ ergs and 
$t_{cav} = 2 \times 10^7$ yrs. 
We also considered several other sizes for the source 
region, $r_s = 0.5$ and 5.0 kpc, but the computed 
flows and our conclusions 
are not sensitive to this parameter. 
At times $t > t_{cav}$ when ${\dot S}_c = 0$, 
the total CR energy $E_c = \int e_c dV$ over the grid 
volume remains very nearly constant at $E_{ctot}$ 
but changes slightly due to advection in adiabatic 
compressions or rarefactions.

We have also considered a range of constant and 
density-dependent CR diffusion coefficients, but 
(unless noted otherwise) the calculations discussed below 
were done with the following variation with 
thermal electron density:
\begin{displaymath}
\kappa = \left\{ 
\begin{array}
{r@{\quad:\quad}l}
10^{30} ~{\rm cm}^2 {\rm s}^{-1} & n_e \le 0.006 ~{\rm cm}^{-3} \\
10^{30}(0.006/n_e) ~{\rm cm}^2 {\rm s}^{-1} & n_e > 0.006 ~{\rm cm}^{-3}
\end{array} \right. 
\end{displaymath}

This $\kappa(n_e)$ is similar to the diffusion coefficients 
used by Mathews \& Brighenti (2007) where a more complete 
discussion can be found. 
We regard $\kappa$ as an adjustable parameter 
although our values are similar to those used in CR 
models of the Milky Way (e.g. Snodin et al. 2006). 
If the cavity jet and outer radio lobe in 
M87/Virgo are related, as we believe, $\kappa$ must be sufficiently 
large so the jet and the lobe evolve on the same time scale. 

\section{A Cosmic Ray plus Gas Dynamical Evolution for M87/Virgo}

Figure 1 shows the X-ray surface brightness 
$\Sigma(r,z)$ and the cosmic ray energy density contours 
of the cavity-lobe evolution at three times. 
At early times (upper and central panels) 
$\Sigma(r,z)$ is the bolometric 
X-ray brightness, but at $10^8$ yr the $\Sigma(r,z)$ is plotted 
in the 
soft energy 0.5 - 1 keV band where the cavity jet 
is illustrated in the {\it Chandra} images of 
Forman et al. (2007).
The bolometric correction for this soft X-ray bandpass 
$BC(T) = \Lambda(T,z_{Fe};0.5-1.0~{\rm keV})/ \Lambda(T,z_{Fe})$ 
was found using XSPEC assuming solar abundances.

The weak shock produced by the growing X-ray cavity is 
faintly visible in the upper and central panels of Figure 1
at 22 and 33 kpc respectively. 
The brightness enhancement at the shock is $\sim20$\%, 
while the emission weighted temperature increases 
by only $\sim5$\%.
The cavity reaches its maximum size at about $2 \times 10^7$ yr 
when the source ${\dot S}_c$ has just turned off.  
A cavity of radius 5 kpc is visible in the central panel 
of Figure 1 centered just slightly beyond the 10 kpc 
CR injection radius.

Most observers estimate the energy required to form 
and fill a cavity of volume $V$ as 
$E_{c,est} = [\gamma_c / (\gamma_c - 1)] P V = 4 P V$ 
where $P$ is the local pressure in the local cluster hot gas. 
At the CR injection radius in Virgo the gas pressure 
is $P = 1.3 \times 10^{-10}$ dynes and 
$E_{c,est} = 8 \times 10^{57}$ ergs 
at $2 \times 10^7$ yr when the cavity radius is 5 kpc. 
However, the CR energy expended at this time in our model 
is $E_{ctot} = 10^{59}$ ergs, about 12 times larger than $4PV$.
 
As the cavity forms in Figure 1, 
the CR pressure contours are nearly concentric 
with the shape of the cavity (upper and central panels). 
At $2 \times 10^7$ yr the cavity jet has not yet been formed, 
although the bottom of the cavity is clearly distorted.
This flattening in the $z$-direction becomes more pronounced
after $2 \times 10^7$ yr and by $5 \times 10^7$ yr 
the crosssectional shape of the buoyant cavity 
(now located at $z \approx 26$ kpc) 
resembles a banana. 
Meanwhile, its overall volume decreases as CRs diffuse 
through the cavity walls. 

The cavity jet becomes clearly visible at $5 \times 10^7$ yr and 
grows in visibility until about $10^8$ yr, seen 
(in the 0.5 - 1 keV bandpass) as a long 
bright extension along the $z$ axis in the 
lower panel of Figure 1. 
The faint vertical feature along the $r$-direction at
$z \approx 33$ kpc may correspond to a similar 
(more one-sided) feature in Fig. 2 of Forman et al. (2007).
We show below that the cavity jet is visible until 
about $2.5 \times 10^8$ yrs.
By $10^8$ yr the CR energy density contours $e_c(r,z)$ are elongated 
along the $z$-axis, resembling the geometry of the large 
outer radio lobe in M87/Virgo.

The detailed variation of gas and CR parameters along the 
$z$ (jet) axis in Figure 1 is illustrated 
in more detail in Figures 2 and 3. 
The top panel of Figure 2 at $2 \times 10^7$ yr 
shows the dominance of the cosmic ray 
pressure in the cavity region, $0.7 \lta \log z_{kpc} \lta 1.37$
($5 \lta z_{kpc} \lta 23$ kpc). 
The abrupt gas pressure change at $\log z_{kpc} = 1.55$
($z_{kpc} = 35$) 
is produced by the weak shock that accompanies
the CR cavity formation. 
After $5 \times 10^7$ yr (central panel) the buoyant cavity 
(with radial diameter $\sim$4 kpc) 
is only slightly visible where $P_c \approx P$
near $\log z_{kpc} \approx 1.4$
($z_{kpc} = 26$ kpc).
Most of the CRs that supported the cavity at $2 \times 10^7$ yr have 
by $5 \times 10^7$ yr diffused into the ambient gas and $P_c(r)$
has flattened considerably. 
At $10^8$ yr the gas pressure in Figure 2 has now nearly 
recovered its pre-cavity profile and the cosmic ray pressure, 
although smaller, extends to 30-40 kpc, comparable to the
size of the M87/Virgo outer radio lobe which has an 
estimated age of $\sim$$10^8$ yr (Owen et al. 2000).
Note that the 
initial undisturbed M87/Virgo atmosphere, shown with the same
dotted profile in each panel of Figure 2, 
differs only very slightly from 
the total pressure $P_{tot} = P + P_c$ (solid lines) 
during most of the evolution. 
This confirms that the gas flow is largely subsonic.

Figure 3 shows the radial gas velocity and density profiles 
along the $z$ (jet) axis at the same three times. 
The density $n_e(z)$ along the cavity jet 
departs significantly from the 
original (observed) profile shown with the same dotted 
line in all three panels. 
At $2 \times 10^7$ yr the density has a sharp minimum that defines the 
moment when the X-ray cavity has reached its 
maximum size, but the cavity center has moved 
somewhat beyond the CR injection radius (10 kpc).
The weak shock is visible at 33 kpc.  
At $2 \times 10^7$ yr the cavity jet has not yet appeared, 
but at $5 \times 10^7$ yr and $10^8$ yr we see a clear density enhancement 
in the cavity jet out to about 22 and 33-40 kpc respectively, 
i.e. $n_e > n_e(t=0)$.
By $5 \times 10^7$ yr the cavity at $\sim$26 kpc has filled in and become 
much smaller. 
By $10^8$ yr no cavity is visible on the $z$ axis. 
But at this time the dense cavity jet 
(dashed line) is even more 
pronounced compared to the original density profile 
(dotted line) and extends to $\sim35$ kpc. 
Higher densities in the cavity jet 
together with the unperturbed 
pressure profile in Figure 2 indicate that the 
cavity jet is much cooler than the surrounding gas, 
consistent with observations of the 
M87/Virgo filament 
(Simionescu et al. 2001; Molendi \& Gastaldello 2001).

The velocity along the $z$ axis, shown in Figure 3 with 
solid lines, is very strongly peaked inside the cavity at $2 \times 10^7$ yr. 
Although the total negative pressure gradient $d(P + P_c)/dz$ 
remains similar to that in the original undisturbed gas, 
the gas density is extremely low inside the cavity 
so the gas acceleration 
$a \approx \rho^{-1}d(P + P_c)/dz \approx \rho^{-1}dP_c/dz$ 
becomes enormous.
High velocities are also visible in the smaller cavity at $5 \times 10^7$ yr. 
Since only a tiny mass is involved, 
these dramatic localized high velocity episodes
have little influence on the overall evolution.

The radial velocity along the cavity jet 
at $5 \times 10^7$ yr and $10^8$ yr 
in Figure 3 describes the dynamics and fate of this 
feature which is so prominent in soft X-ray images 
of M87/Virgo.
At $5 \times 10^7$ yr the jet velocity is positive beyond 8 kpc 
and increases almost linearly to nearly 600 km s$^{-1}$ 
at 20 kpc where the sound speed in the original 
atmosphere is 720 km s$^{-1}$. 
However, by $10^8$ yr the jet velocity along the 
cavity jet is negative within about 22 kpc  
and is falling back toward the M87/Virgo core. 
The maximum velocity 200 km s$^{-1}$ at 30 kpc 
is now lower. 
As we show below, by $\sim 3 \times 10^8$ yrs the 
entire cavity jet has fallen back and shocked with the 
dense gas at the center of M87. 
(It is possible that the infalling cavity jet 
contributed to the black hole accretion that triggered 
the energy outburst resulting in the younger 
non-thermal jet in M87.)

Figure 4 provides a snapshot of the variation of 
gas properties perpendicular to the cavity jet 
at $z = 15$ kpc and time $10^8$ yr.
The solid lines in these plots are profiles in the 
$r$-direction across the jet at $z = 15$ kpc. 
For comparison, the dashed lines 
are profiles in the gas (far from the jet) 
along the $z$-direction at $r = 15$ kpc.
The deviation of the solid profiles from the dashed profiles 
shows the local variations caused by the cavity jet.  
In the top two panels the density and temperature 
(light solid lines) deviate from the undisturbed gas 
in opposite senses so that the pressure across the jet 
(third panel down) is essentially unchanged. 
Similar snapshot figures at other times and $z$ distances 
reveal that the cavity jet is in pressure equilibrium 
with the surrounding gas throughout most of its evolution.
The emission-weighted temperature profiles, shown with heavy lines 
in the second panel, confirm that the colder jet is still 
clearly visible when observed through the entire 
cluster gas. 

The low entropy $S = kT/n_e^{2/3}$ and high (iron) abundance 
in the profiles transverse to the jet shown in Figure 4 
attest that the gas in the cavity jet has 
come from deep within the core of M87/Virgo. 
As observers search for similar radial cavity jets in other 
clusters, they should appear as radial features of 
low temperature, low entropy and high abundance. 
These are all well-known attributes of the 
cavity jet in M87/Virgo.

The velocity structure transverse to the cavity jet 
$v_z(z=15~{\rm kpc},r)$ 
shown in Figure 4 reveals that the core of the jet 
at $z = 15$ kpc and time $10^8$ yr is already falling back toward 
the M87 core while most of the gas beyond about 1.5 kpc 
from the jet center is still moving out.
At earlier times, the jet center can also have positive velocity, but 
it always moves more slowly than the periphery of the jet
throughout the evolution. 
(Note that the core velocities 
shown in Figure 3 may not represent the mean velocity 
of most of the jet gas at that radius.) 
The cores of the cavity jet reach their maximum 
radius and begin to fall back somewhat sooner than 
the outer (more massive) parts.
This behavior is expected because at any time the 
cavity jet is in pressure 
equilibrium with the ambient gas at each radius and 
also shares the same gradient $dP_{tot}/dz$ along the jet.
Consequently, the outward acceleration of gas in the jet 
$a \approx \rho^{-1}dP_{tot}/dz$ 
is therefore always less in the core of the jet 
where the density is greatest.

Figure 5 shows X-ray surface brightness profiles 
$\Sigma(z,r)$ in the 
soft 0.5 - 1.0 keV band similar to that illustrated 
in Fig. 4 of 
Forman et al. (2007) for the M87/Virgo filament.
The solid lines show $\Sigma(r)$ at fixed values 
of $z$ that label the curves; 
these illustrate the decrease in X-ray surface brightness 
proceeding from the center of the cavity jet 
in a perpendicular direction. 
For comparison, 
the dashed lines show $\Sigma(z)$ far from the jet at the same 
radius $r$ labeled in each panel.  
As time progresses the half-width of the computed 
filament in Figure 5 slowly shrinks until, at 
$3 \times 10^8$ years the entire filament 
(and most of its iron) has fallen back 
to the core of M87. 
The observed surface brightness enhancement in M87/Virgo 
at 10 kpc along the jet is 
(shown in Fig. 4 of Forman et al.)
has a maximum contrast to 
the local gas of about 20 - 30 percent
(or 0.08 - 0.12 in $\log \Sigma$) and the 
half-width is 5-10$^{\prime\prime}$ (0.4 - 0.8 kpc).
Our calculated cavity jet (filament) has similar properties but 
is somewhat broader and more luminous than that in 
M87/Virgo.
The occasional multiplicity of peaks 
in Figure 5 also appears in 
the M87/Virgo observations, but such irregularities 
would be expected in the more active environment of a 
real cluster.
As a check, we repeated the cavity jet calculation 
at higher resolution using a 
numerical grid about three times smaller and 
the features shown in Figures 4 and 5 
were essentially unchanged. 


Cooling by radiation losses in equation 3 does not strongly influence
our results.  
In some of our cavity jet calculations we found that at
time $10^8$ yrs the radiative cooling time in the cavity jet $t_{cool}
= 5 m_p k T/ 2 \mu \rho \Lambda$ was also $\sim10^8$ yrs.  
However, as
the calculation proceeded, adiabatic cooling continued to dominate and
the radiative cooling time $t_{cool}$ eventually increased faster than
the time elapsed.  
Although $t_{cool}$ in the cavity jet can pass
through a minimum, no gas cooled to very low temperatures.  
It is
possible that radiative cooling to low temperatures may occur with
somewhat different cluster and cavity parameters.  
In addition, after
a few $10^8$ yrs a small mass of gas cooled near the origin, but this
has no effect on the results discussed here.

Figure 6 illustrates rather dramatically 
the (temporary) transport of low entropy, metal-rich 
gas in the cavity jet from the central core to 
large distances in the hot cluster gas at $10^8$ yr. 
The entropy and abundance along the filament core 
are both remarkably constant out to 30-40 kpc, 
showing little effects due to numerical diffusion.
This near constancy of the entropy is possible only
if the filament gas expands adiabatically without 
experiencing shocks.

\section{Radio Synchrotron Emission in M87/Virgo 
Observed and Estimated}

We now inquire if the cosmic ray energy density $e_c(r,z)$ 
shown at $10^8$ yr in Figures 1 and 2 is sufficient to emit the 
radio luminosity observed in the M87/Virgo halo.
The half life for relativistic electrons to synchrotron 
losses is $t_{1/2} \approx 25 /B^2 \gamma_L$ yrs, 
where for simplicity the Lorentz factor $\gamma_L$ is 
assumed to be the same for all electrons.
Taking the mean field in M87/Virgo to be $B = 10\mu$G 
(e.g. Owen et al. 2000) 
and imposing $t_{1/2} = 10^8$ yrs, 
we find $\gamma_L \approx 2500$.
The total synchrotron power from each electron is 
$P_{syn} = (4/3)\gamma_L^2 c \sigma_T (B^2 / 8\pi) 
= 7 \times 10^{-19}$ erg s$^{-1}$. 
The mean number density of relativistic electrons $n_c$
can be estimated using 
$n_c \gamma_L m_e c^2 = e_c = 3P_c \approx 3 \times 10^{-11}$ 
erg cm$^{-3}$, taken from the $t = 10^8$ year panel in Figure 2. 
It follows that $n_c \approx 1.5 \times 10^{-8}$ cm$^{-3}$ 
so the total emissivity from all electrons is 
$n_c P_{syn} \approx 10^{-26}$ erg cm$^{-3}$ s$^{-1}$.
Then we estimate 
the total radio luminosity from the radio lobe
(of radius $r_{lobe} \approx 35$ kpc) to be 
$L \approx n_cP_{syn}(4/3) \pi r_{lobe}^3 \approx 5 \times 10^{43}$ 
erg s$^{-1}$. 

The total radio luminosity observed in M87/Virgo 
from 10MHz to 150GHz is $9 \times 10^{41}$ ergs s$^{-1}$ 
(Herbig \& Readhead 1992).
However, we estimate the luminosity of the radio halo alone, 
$1.5 \times 10^{41}$ erg s$^{-1}$, 
by integrating the halo flux densities tabulated 
by Andernach et al. (1979) from 0.1 to 10GHz.

We see that the observed radio luminosity is about 300  
times less than our rough idealized estimate. 
This suggests that a more detailed radio 
synchrotron model consisting 
of a range of relativistic electron energies and 
including relativistic protons 
can be designed to fit the radio data. 
We do not attempt such a detailed model at present since 
it introduces additional sensitive, poorly known parameters, 
such as the radial dependence of the magnetic field.

One possible concern with our treatment of CR diffusion is that the
dimensions of the large radio halo in M87/Virgo 
would be similar at all radio frequencies. 
If more energetic electrons diffuse appreciably faster,
as might be expected, the size of the radio halo
could vary with frequency (Owen et al. 2000).
However, the energy estimates above suggest that 
a wide energy spectrum of secondary electrons can be produced 
throughout the outer lobe by pion production from cosmic ray protons. 
Owen et al. (2000) also note that the outer boundary of the
radio halo appears sharper than might be predicted by 
a standard diffusion model. 
However, the physical nature of the CR diffusion coefficient
is still largely unknown and is likely to differ from the vanilla 
diffusion considered here.  
To better understand cosmic ray diffusion, we urge radio 
observers to determine the precise shape of the ``sharp'' 
outer boundary of the outer lobes in M87/Virgo at several 
radio frequencies. 

The shape and size of the cosmic ray lobe depends sensitively 
on the diffusion coefficient $\kappa(n_e)$. 
We illustrate one example of this in  
Figure 7 which shows athe CR contours (and soft X-ray image) at 
time $10^8$ yr in a calculation similar to that shown 
in Figure 1 but different in two respects: (1) the 
total CR energy is reduced to 
$E_{ctot} = 5 \times 10^{58}$ ergs to keep the 
cavity size at $2 \times 10^7$ yrs similar to that in Figure 1, and 
(2) the CR diffusion coefficient is lowered as follows: 

\begin{displaymath}
\kappa = \left\{
\begin{array}
{r@{\quad:\quad}l}
10^{29} ~{\rm cm}^2 {\rm s}^{-1} & n_e \le 0.008 ~{\rm cm}^{-3} \\
10^{29}(0.008/n_e)^{1.14} ~{\rm cm}^2 {\rm s}^{-1} 
& n_e > 0.008 ~{\rm  cm}^{-3}
\end{array} \right.
\end{displaymath}
With this lower $\kappa(n_e)$, the CR contours 
(if symmetric about the $z = 0$ plane)  
more clearly define two separate cosmic ray (outer radio) 
lobes that have sharper edges. 
The thermal cavity jet is comparable in brightness and other 
properties to the one we have described in detail above. 

\section{Cavity Jets and ``Relic'' Radio Sources 
in Other Clusters}

Many galaxy clusters contain 
extended, steep spectrum ``relic'' radio source located far 
from the cluster-centered galaxy but which are unassociated with any 
line of sight galaxy near the source  
(Giovannini \& Feretti 2004).  
The dimensions and locations of relic sources vary considerably, 
but most relics are extremely large ($\gta 500$ kpc) and very 
distant ($\sim 1000$ kpc) from the cluster center. 
One much-discussed origin for giant relic sources are merger-related
shocks that re-energize latent cosmic ray electrons in the cluster
gas (e.g. Pfrommer, Ensslin, Springel, 2007).
However, there is a subclass of smaller relics -- like 
those in A13, A85, A133, A4038 (Feretti \& Giovannini 2007) -- 
that often lie within or near X-ray detectable cluster gas, 
and are therefore relevant to our calculations here.
To our knowledge the outer radio lobes of M87/Virgo 
have not been classified as relic sources, but they 
resemble relics in many ways except 
we view them projected against the cluster-centered galaxy M87.
The outer lobes in M87/Virgo may be younger than most relics.
If the magnetic field decreases 
with cluster radius (and CR electron lifetimes increase), 
radio electrons closer to the cluster center will be lost and 
the mean radius of the M87/Virgo lobes will 
increase with time. 

The cluster A133 contains a relic radio source 
(Slee et al. 2001) located about 40 kpc NW of the cluster center 
and {\it Chandra} observations show an extended filament of 
relatively cool gas proceeding radially from the cluster core 
to the relic source (Fujita et al. 2002). 
This is the same radio and X-ray configuration 
we show in Figures 1 (bottom panel) and 7. 
The mean gas temperature in A133 rises from 2 keV at a few kpc 
to 4.5 keV at 110 kpc, but the temperature of the dense thermal 
filament is lower (e.g. less than 2 keV at $\sim20$ kpc). 
X-ray and radio observations of A13,
another similar but perhaps more complicated cluster, 
have been discussed by Juett et al. (2007). 
In A13 a broad radial filament of cooler gas connects the cluster 
core to the relic radio source at $\sim300$ kpc.
The authors of these papers suggest that while the radio emission 
probably comes from a central AGN, the cool filaments 
either result from a recent merger or 
from an ``uplifting'' of cooler gas from the cluster core. 
We prefer to think that radio relics and radial X-ray filaments 
are both natural outcomes of cavities formed by cosmic rays.

\section{Concluding Remarks}

Radial jet-like thermal filaments 
and associated large radio lobes evolve from X-ray 
cavities that are inflated with cosmic rays 
that eventually diffuse through the cavity walls. 
The radio features created from cavity CRs 
are unlike either FRI or FRII morphologies, 
where non-thermal energy flows 
directly from the central black hole. 
In general cavities and cavity jets are not likely 
to coexist.
X-ray cavities containing cosmic rays that later diffuse into larger
radio lobes are likely to be observed early-on as they are being
filled with cosmic rays or shortly thereafter.
By the time the cosmic rays have diffused from an 
X-ray cavity to fill a large radio-emitting volume in the 
cluster gas, the initial cavity will have disappeared. 
The radio lobes will be accompanied 
for $\gta 2 \times 10^8$ yrs 
by nearly radial,
low temperature, low entropy, metal rich thermal filaments 
similar to the M87/Virgo filament seen in {\it Chandra} images. 

Since X-ray cavities are observed in about 20 percent 
of all clusters (Birzan et al. 2004), 
it may be surprising that the fraction of clusters 
like M87/Virgo, A13 and A133 with cavity jets is much smaller. 
Moreover, 
the double outer radio lobes in M87/Virgo appear to be 
nearly identical radio sources symmetric about 
the center of M87, but only one lobe has a cavity jet.
If the filaments last about 4-5 times longer than 
X-ray cavities, as in our model, 
the number of cluster filaments should be larger 
by the same factor, but this has not been observed to date. 

The initial position of the cavity does not seem to have 
a large effect on the development or X-ray visibility 
of cooler cavity jets over times $\sim10^8$ yrs.
To explore this we repeated the calculation 
described in Figure 7 but with the initial cavity 
formed at $z = 20$ kpc instead of 10 kpc. 
In this case the cavity jet was broader and remained visible 
for slightly longer times
(but with a somewhat lower amplitude). 
The cavity jet is visible from well within 20 kpc 
from the center of M87/Virgo to 30 kpc and beyond.
The CR energy density $e_c(r,z)$ at time $10^8$ yrs
resembles that in Figure 7 but with contours shifted out 
by $\sim 10$ kpc along the $z$-axis. 

Evidently many cluster filaments have been missed because 
most X-ray clusters are much fainter than M87/Virgo 
and have not been observed with deep, 
$\sim$1000 ksec {\it Chandra} exposures. 
If the CR diffusion coefficient $\kappa$ is larger 
than our values for M87/Virgo, the cavities will be smaller 
for given $E_{ctot}$ 
and their lifetimes may be too short to form cavity jets. 
Nevertheless, we hope that our calculations will inspire 
a more dedicated search for additional cavity jets 
in X-ray cooling core clusters having large diffuse radio lobes. 
Combined observations of thermal filaments and diffuse radio 
lobes can be used to date the explosive events from which 
they both evolved and to study in detail the propagation 
of cosmic rays, the magnetic structure in the hot gas 
and the total cosmic ray energy involved. 
If M87/Virgo is typical, the total energy of cosmic rays 
released $E_{ctot}$ can exceed the work necessary 
to simply form and fill the cavities, 
$E_{c,est} = 4 P V$, by more than an 
order of magnitude. 

The hot gas in the Virgo cluster is currently experiencing 
a sequence of strong -- and very different -- energy 
releases from the central black hole.
In addition to the cavity jet event discussed here, 
the more famous 
(and more recent) 2 kpc jet is moving in a NW direction 
unrelated to the orientation of the 
cavity jet and its large radio lobes. 
In addition, 
a strongly buoyant mushroom-shaped flow directly to the East of 
M87 is seen in both X-rays and radio images 
(Churazov et al. 2001). 
Ruszkowski et al. (2007) propose that an intensified 
magnetic field in the buoyant stem of this flow has caused
it to be prominent in the 90 cm radio image 
of Owen et al. (2000).  
However, the cavity jet in the {\it Chandra} 
image of M87/Virgo does not appear in this radio image. 
This difference in the magnetic field strength 
may result because the gas in the Eastern plume 
originated deeper within the cluster gas than that 
in the cavity jet, 
assuming that the field strength increases 
(with gas density) toward the cluster core. 

In addition to these three energy outflows associated 
with strong radio emission, 
a number of less energetic outflows from the core of M87 
are apparent in optical emission. 
Fig. 2 of Young et al. (2002) 
shows the central 7 $\times$ 7 kpc region of M87/Virgo 
at three very different frequencies: {\it Chandra} 
X-ray, 6 cm radio and optical emission in 
the H$\alpha$ + [NII] lines.
The optical line emission is particularly interesting 
because about 5 plumes of warm ($T \sim 10^4$ K) gas 
can be seen emanating from the M87 core, 
all in different directions. 
The brightest of these is a H$\alpha$ + [NII] feature 
that lies just North of the bright 2 kpc 
non-thermal jet, but at a slightly different angle. 
We see a parallel here with the dusty plume we have 
discovered and discussed in the X-ray bright galaxy 
group NGC 5044 (Temi, Brighenti \& Mathews 2007).
The NGC 5044 feature can be understood as dusty buoyant gas heated 
near the central black hole that is rapidly cooled to 
$\sim10^4$ K by electron-dust cooling. 
It seems entirely plausible that the warm gas 
H$\alpha$ + [NII] features 
in M87 are created in the same manner, adding to the 
already wide diversity of energy releases in this 
remarkable galaxy/cluster.

We expect that the numerous intermittent energy 
events seen in M87/Virgo have created subsonic velocities 
throughout the hot gaseous atmosphere. 
By contrast, our cavity jet was computed starting 
with stationary cluster gas exactly in hydrodynamic equilibrium 
and this results in a perfectly axisymmetric flow 
that may differ in detail from that observed in M87/Virgo. 
For example,
the slight curvature of the cavity jet visible in 
Fig. 2 of Forman et al. (2007) 
could arise from either from a small velocity perpendicular 
to the accelerating jet or from 
an initial velocity asymmetry near the base of the pre-filament 
cavity that slightly altered the direction of the cavity jet 
as it formed. 
Similarly, the ``braided'' nature of this filament 
seen in the {\it Chandra} image 
(Fig. 3 in Forman et al. 2007) may result from small 
random transverse velocities 
present when the X-ray cavity was formed. 

\vskip0.1in
Studies of the evolution of hot gas in elliptical galaxies
at UC Santa Cruz are supported by
NASA grants NAG 5-8409 \& ATP02-0122-0079 and NSF grant
AST-0098351 for which we are very grateful.


\clearpage
\begin{figure}
\centering
\vskip1.in
\includegraphics[bb=90 166 522 519,scale=0.8,angle= 0]
{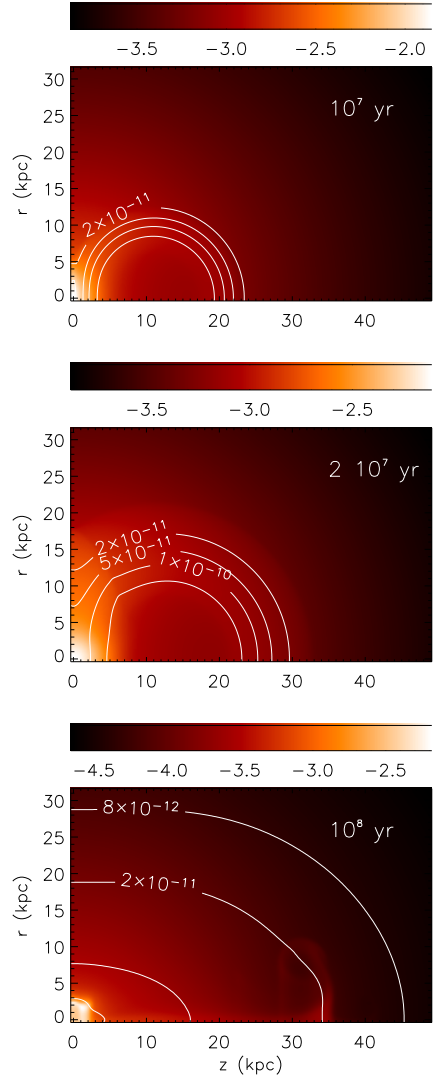}
\vskip1.in
\caption{
X-ray surface brightness images from computed models
at three times: $10^7$, $2 \times 10^7$ and $10^8$ yrs.
The cavity and jet are formed on the horizontal ($z$) axis. 
For the first two times 
the bolometic surface brightness $\Sigma(r,z)$
(in ergs cm$^{-2}$ s$^{-1}$) is shown and calibrated 
in the color bars (upper two panels). 
At $10^8$ years (lower panel)
the surface brightness is restricted to
the 0.5 - 1 keV bandpass to reveal the colder jet filament 
along the $z$ axis.
Each plot shows the following contours of the cosmic ray
energy density $e_c(r,z)$ (from the inside out):
at $10^7$ yrs: $2 \times 10^{-11}$, $5 \times 10^{-11}$,
$10^{-10}$, and $2 \times 10^{-10}$;
at $2 \times 10^7$ yrs: $2 \times 10^{-11}$, $5 \times 10^{-11}$,
$10^{-10}$, and $2 \times 10^{-10}$;
at $10^8$ yrs: $8 \times 10^{-12}$, $2 \times 10^{-11}$, 
$5 \times 10^{-11}$, and $8 \times 10^{-11}$,
all in ergs cm$^{-3}$.
}
\label{f1}
\end{figure}

\clearpage
\begin{figure}
\vskip2.in
\centering
\includegraphics[bb=90 166 522 519,scale=0.8,angle= 0]
{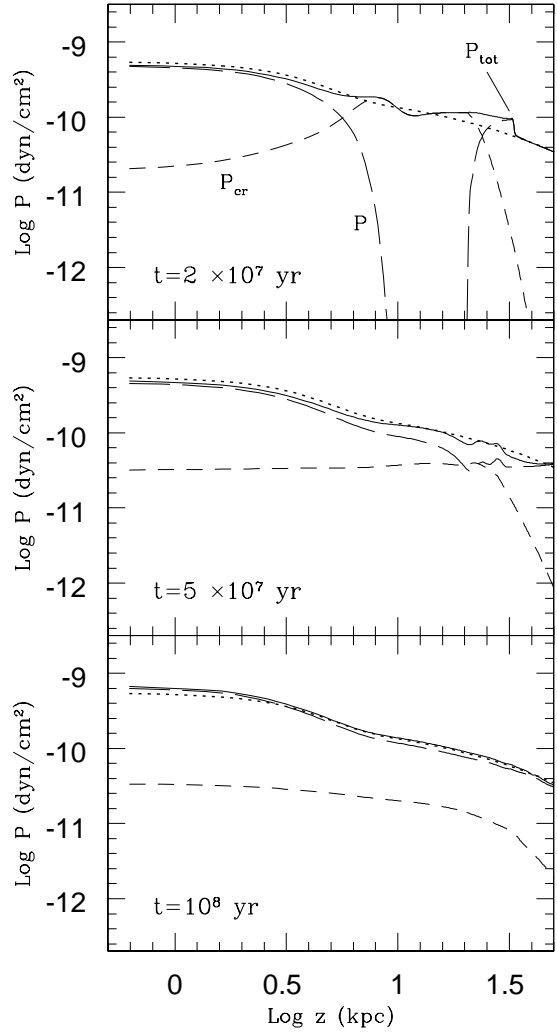}
\vskip.7in
\caption{
Variation of pressures 
along the center of the cavity jet ($z$ axis) at
three times: $2 \times 10^7$ yrs (top panel), 
$5 \times 10^7$ yrs (middle panel), and 
$10^8$ yrs (bottom panel). 
Each panel shows the gas pressure $P$ (long dashed lines), 
cosmic ray pressure $P_c$ (short dashed lines),  
total pressure $P + P_c$ (solid lines).
Also shown is the gas pressure profile 
in the initial undisturbed M87/Virgo 
atmosphere before the cavity was created (dotted lines).
}
\label{f2}
\end{figure}

\clearpage
\begin{figure}
\vskip2.in
\centering
\includegraphics[bb=90 166 522 519,scale=0.8,angle= 0]
{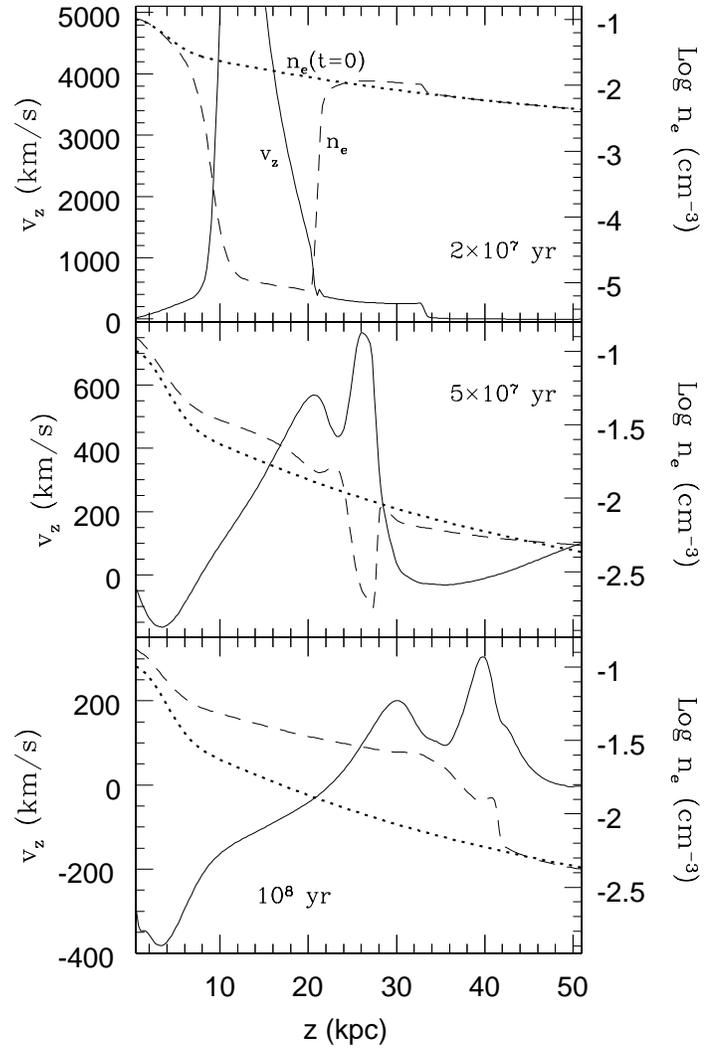}
\vskip.7in
\caption{
Variation of electron density 
$n_e$ (dashed lines) and gas velocity $v_z$ (solid lines)  
along the center of the cavity jet ($z$ axis) at 
three times. 
Also shown is the electron density $n_e$ profile  
in the initial undisturbed M87/Virgo atmosphere (dotted lines).
}
\label{f3}
\end{figure}

\clearpage
\begin{figure}
\vskip2.in
\centering
\includegraphics[bb=90 166 522 519,scale=0.8,angle= 0]
{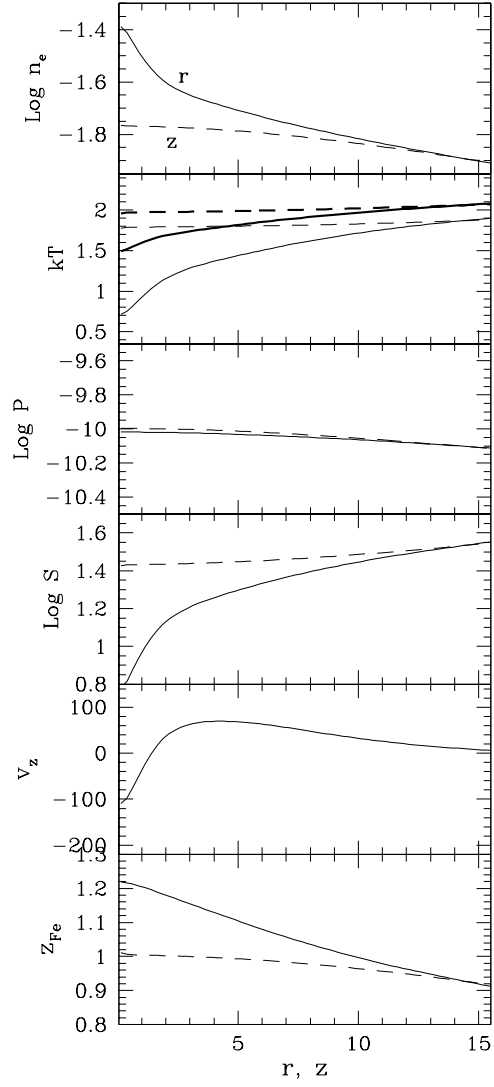}
\vskip.7in
\caption{
Light solid lines show profiles in the $r$-direction 
of the electron density $n_e$ (cm$^{-3}$), 
$kT$ (keV), gas pressure $P$ (dynes), entropy 
$S = kT/n_e^{2/3}$ (keV cm$^{-2}$), velocity $v_z$ 
(km s$^{-1}$), and iron abundance $z_{Fe}$ (solar units) 
perpendicular to the cavity jet at $z = 15$ kpc 
at time $10^8$ yrs. 
For comparison the 
dashed lines show the profiles in the $z$-direction 
at $r = 15$ kpc at the same time far from the jet. 
The heavy solid and dashed lines show the variation of the 
emission-weighted temperature when viewed through the 
entire M87/Virgo atmosphere. 
}
\label{f4}
\end{figure}

\clearpage
\begin{figure}
\vskip2.in
\centering
\includegraphics[bb=90 166 522 519,scale=0.8,angle= 0]
{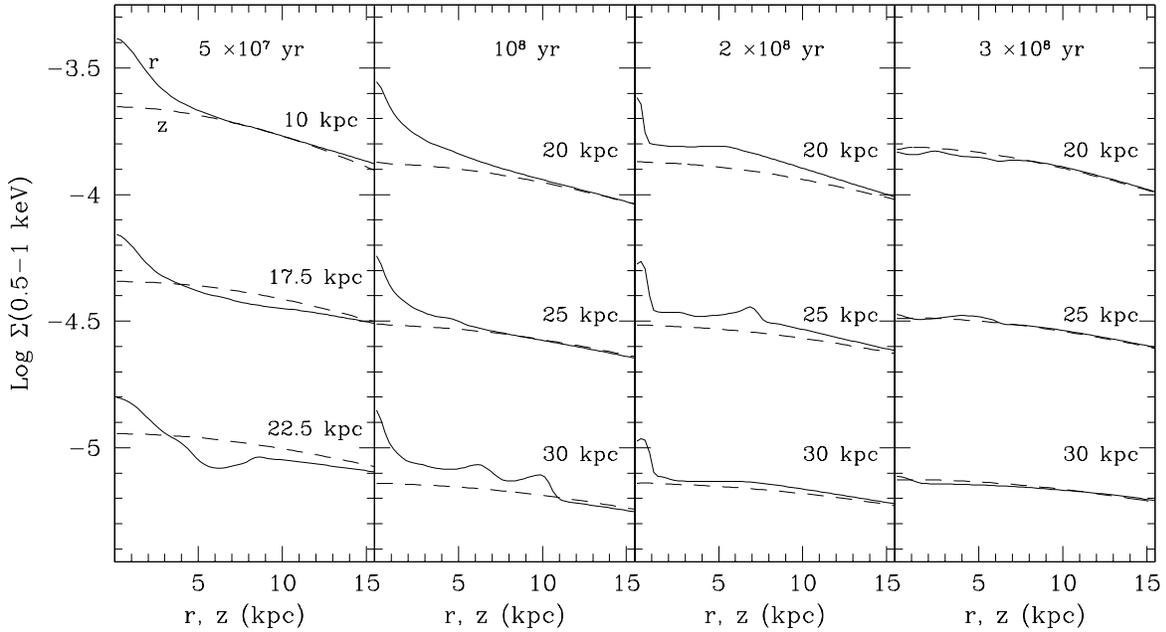}
\vskip.7in
\caption{
Profiles at four times 
of the soft X-ray surface brightness of M87/Virgo 
$\Sigma(r,z)$ (0.5 - 1 keV) perpendicular to 
the cavity jet along the $z$ axis 
compared with $\Sigma(r,z)$ along the $r$ axis far from the jet.
Solid lines show the variation of $\Sigma$ 
(erg cm$^{-2}$ s$^{-1}$) in the 
$r$-direction perpendicular to the jet at distances 
$z$ along the jet that label each curve. 
Dashed lines show the variation of $\Sigma$ in the
$z$-direction perpendicular to M87/Virgo 
(far from the jet) at distances
$r$ along the $r$-axis that label each curve.
For visibility 
each of the three profile pairs in the panels has been 
arbitrarily shifted vertically so the scale is only relative.
}
\label{f5}
\end{figure}

\clearpage
\begin{figure}
\vskip2.in
\centering
\includegraphics[bb=90 166 522 519,scale=0.8,angle= 0]
{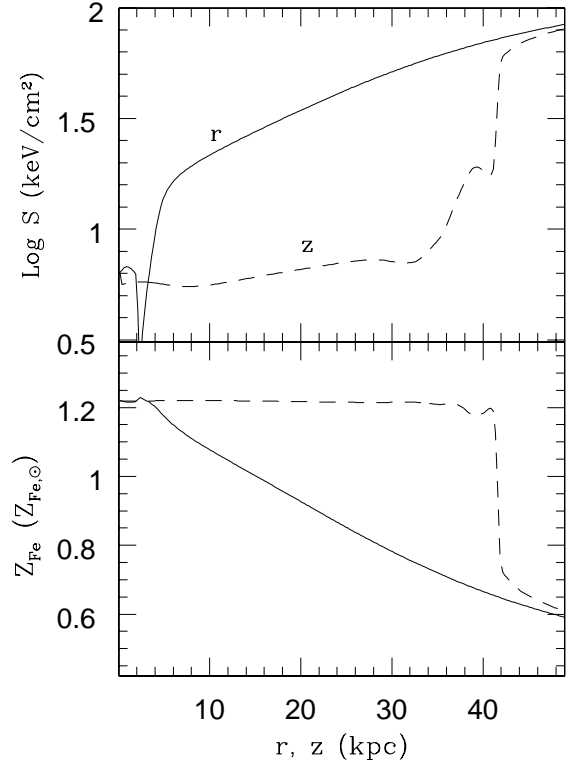}
\vskip.7in
\caption{
Radial profiles at time $10^8$ yrs 
of the entropy $S = kT/n_e^{2/3}$ (keV cm$^{-2}$) 
and iron abundance $z_{Fe}$ (solar units) directly along the 
jet ($z$-axis) shown with dashed lines 
and far from the jet ($r$-axis) shown with solid lines.
}
\label{f6}
\end{figure}

\clearpage
\begin{figure}
\vskip2.in
\centering
\includegraphics[bb=90 166 522 519,scale=0.8,angle= 0]
{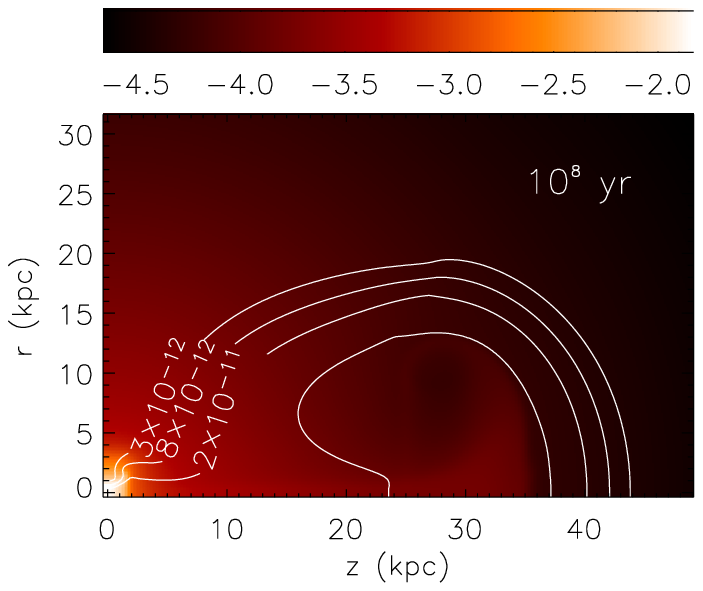}
\vskip.7in
\caption{
X-ray surface brightness $\Sigma(r,z)$ (erg cm$^{-2}$ s$^{-1}$) 
in the 0.5 - 1 keV bandpass at time $10^8$ yrs 
for a calculation identical to that shown in the lower panel 
of Figure 1 except a with somewhat lower cosmic ray diffusion 
coefficient (see \S 4) and a lower $E_{ctot} = 5 \times 10^{58}$ 
ergs. 
The cosmic ray energy density $e_c(r,z)$ is shown with contours 
(from the outside in) at $3 \times 10^{-12}$, $8 \times 10^{-12}$, 
$2 \times 10^{-11}$, and $6 \times 10^{-11}$, 
all in ergs cm$^{-3}$.
}
\label{f7}
\end{figure}


\begin{references}

\reference{} Andernach, H., Baker, J. R., von Kap-herr, A. \&
Wielebinski, R. 1979, A\&A, 74, 93

\reference{} Birzan, L., McNamara, B. R., Carilli, L., 
Nulsen, P. E. J. \& Wise, M. W. 2006, 
(arXiv:astro-ph/0612393)

\reference{} Birzan, L., Rafferty, D. A., McNamara, B. R., et al.
2004, ApJ, 607, 800

\reference{} Bohringer, H., Voges, W., Fabian, A. C. et al. 
1993, MNRAS, 264, L25

\reference{} Churazov, E., Bruggen, M., Kaiser, C. R., et al.
2001, ApJ, 554, 261

\reference{} Clarke, T., Blanton, E.,  Sarazin, C., et al. 
2006 (arXiv:astro-ph/0612595)

\reference{} Drury, L. O. \& Falle, S. A. E. G., 1986,
MNRAS, 223, 353

\reference{} Fabian, A. C., Sanders, J. S., Ettori, S. 
2000, MNRAS, 318, L65

\reference{} Feretti, L. \& Giovannini, G. 2007,  
to be published in Springer Lecture Notes in Physics, ``Panchromatic
View of Clusters of Galaxies and the Large-Scale Structure'', Editors
M. Plionis, O. Lopez-Cruz, and D. Hughes (arXiv:astro-ph/0703494)

\reference{} Forman, W., Jones, C., Churazov, E. et al.
2007, ApJ, 665, 1057

\reference{} Forman, W., Nulsen, P., Heinz, S., et al.
2005, ApJ, 635, 894

\reference{} Fujita, Y., Sarazin, C. L., Kempner, J. C., et al. 
2002, ApJ, 575, 764

\reference{} Gardini, A. 2007, A\&A, 464, 143

\reference{} Giovannini, G. \& Feretti, L.  2004, Korean Astron. Soc., 
37, 323

\reference{} Govoni, F. \& Feretti, L. 2004, Intenational Journal of
Modern Physics D, 13, 1549

\reference{} Ghizzardi, S., Molendi, S., Pizzolato, F. \&
De Grandi, S. 2004, ApJ, 609, 638

\reference{} Herbig, T. \& Readhead, A. C. S. 1992, ApJS, 81, 83

\reference{} Jones, T. W. \& Kang, W. 1990, ApJ, 363, 499

\reference{} Juett, A. M., Sarazin, C. L., Clarke, T. E., et al. 
2007, ApJ (in press) (astro-ph/arXiv:0708.2277)

\reference{} Lyutikov, M. 2007, ApJ, (submitted) 
(arXiv:0709.1712)

\reference{} Mathews, W. G. \& Brighenti, F. 2007,
ApJ, 660, 1137

\reference {} Molendi, S. 2002, ApJ, 580, 815

\reference{} Molendi, S.,  \& Gastaldello, F. 2001, A\&A, 375, L14

\reference{} Owen, F. N., Eilek, J. A., \& Kassim, N. E.
2000, ApJ, 543, 611

\reference{} Pfrommer, C., Ensslin, T. A., \& Springel, V. 2007, 
MNRAS (submitted) (arXiv:astro-ph/0707.1707)

\reference{} Rebusco, P., Churazov, E., Bohringer, H. \&
Forman, W. 2006, MNRAS, 372, 1840

\reference{} Reynolds, C. S., McKernan, B., Fabian, A. C. et al.
2005, MNRAS 357, 242

\reference{} Roediger, E., Br\"uggen, M., Rebusco, P., 
B\"ohringer, H. \& Churazov, E. 2007,MNRAS, 375, 15

\reference{} Ruszkowski, M., Ensslin, T. A., Bruggen, M. et al.
2007, MNRAS, 378, 662

\reference{} Simionescu, A., Bohringer, H., Bruggen, M., 
\& Finoguenov, A., 201, A\&A, 465, 749

\reference{} Slee, O. B., Roy, A. L., Murgia, M. et al. 2001,
AJ, 122, 1172

\reference{} Snodin, A. P., Brandenburg, A., Mee, A. J. \& Shukurov,
A. 2006, MNRAS, 373, 643

\reference{} Stone, J. M. \& Norman, M. L. 1992,
ApJS, 80, 753

\reference{} Sutherland, R. S. \& Dopita, M. A. 1993, 
ApJS, 88, 25

\reference{} Temi, P., Brighenti, F., \& Mathews, W. G. 2007,
ApJ, 666, 222

\reference{} Tonry, J. et al. 2001, ApJ, 546, 681

\reference{} Young, A. J., Wilson, A. S. \& Mundell, C. G.
2002, ApJ, 579. 560

\end{references}
\end{document}